\documentclass[aip,reprint]{revtex4-1}

\usepackage{graphicx}

\usepackage{amssymb}
\usepackage{amsmath} 

\begin{document}


\title{Analytical foundation for adversarial synchronization control in oscillator networks} 



\author{Kazuhiro Takemoto}%
\email{takemoto.kazuhiro035@m.kyutech.ac.jp}
\affiliation{
Department of Bioscience and Bioinformatics, Kyushu Institute of Technology, Iizuka, Fukuoka, Japan
}%
\affiliation{
Data Science and AI Research Center, Kyushu Institute of Technology, Iizuka, Fukuoka, Japan
}%


\date{\today}

\begin{abstract}
This study provides an analytical foundation for adversarial synchronization control in Kuramoto oscillator networks, where small gradient-based perturbations applied repeatedly to oscillator phases can dramatically enhance or suppress collective synchronization. Using the Ott--Antonsen reduction, we derive an exact closed-form expression for the effect of a single adversarial perturbation (kick) on the order parameter. A key finding is that each kick produces a finite, coupling-independent increment in the order parameter even when synchronization is arbitrarily weak, which combined with slow relaxation near the critical coupling and mean-field feedback explains the disproportionate amplification previously observed in numerical simulations. Fixed-point analysis further reveals a fundamental asymmetry between enhancement and suppression, with the latter governed by noise-induced escape in finite systems. Extending the framework to networks via the annealed network approximation, we show that the theory captures the synchronization behavior of representative model networks and identify a decoupling between kick sensitivity and mean-field dominance in scale-free networks. These results offer a tractable theoretical basis for understanding and designing kick-based synchronization control in oscillator networks.
\end{abstract}


\maketitle 

\begin{quotation}
Synchronization, the spontaneous alignment of rhythms across many 
coupled units, underlies the stable operation of power grids, the 
coordination of neurons in the brain, and collective behavior in 
biological and social systems. A recently proposed strategy inspired 
by adversarial attack principles from artificial intelligence showed 
that extremely small repeated perturbations can dramatically enhance 
or suppress synchronization across entire networks, but the 
theoretical reason for this outsized effect was not understood. 
This study provides an analytical explanation using an exact 
mathematical framework. We derive a closed-form formula that 
quantifies how each perturbation shifts the degree of 
synchronization, and identify the key factors that amplify a tiny 
intervention into a large-scale effect. We also reveal a fundamental 
asymmetry: enhancing synchronization is robust, while suppressing 
it is limited by finite-size effects. These results provide a 
theoretical basis for designing minimally invasive synchronization 
control in engineered and biological networks.
\end{quotation}

\section{Introduction}
Synchronization of coupled oscillators is a pervasive phenomenon 
in both natural and engineered systems, with prominent examples 
ranging from neuronal assemblies and circadian clocks to power 
transmission networks \cite{arenas2008synchronization,suykens2008introduction,dorfler2014synchronization}.
The Kuramoto model provides a canonical framework for studying 
such collective dynamics \cite{acebron2005kuramoto,rodrigues2016kuramoto}, 
and has motivated a wide range of synchronization control strategies, 
including topology modification, natural frequency manipulation, 
coupling weight adjustment, and external intervention 
\cite{zhou2006dynamical,wang2002pinning,rosenblum2004delayed,menara2022functional}.

Recently, \citet{nagahama2025adversarial} introduced an alternative control 
strategy inspired by adversarial attacks in deep learning 
\cite{yuan2019adversarial,chakraborty2021survey}: small gradient-based 
perturbations applied repeatedly to oscillator phases can dramatically 
enhance or suppress synchronization in Kuramoto networks, exploiting the 
system's intrinsic sensitivities with minimal intervention. Through 
extensive numerical simulations on model and real-world networks, they 
demonstrated that this adversarial framework achieves effective 
synchronization control far beyond what the perturbation magnitude alone 
would suggest. However, the theoretical mechanism underlying this 
disproportionate amplification has remained unexplained.

In this paper, we provide the first analytical foundation for these 
observations. The Ott--Antonsen (OA) reduction \cite{ott2008low,ott2009long} 
has proven to be a powerful and versatile framework for the exact 
analysis of coupled phase oscillator systems, with successful applications to synchronization transitions, 
chimera states, externally forced oscillator populations, and 
network synchronization dynamics 
\cite{abrams2008solvable,pietras2016ott,laing2009chimera,yoon2015critical}. 
For Lorentzian frequency distributions, it reduces the dynamics of 
the infinite-dimensional phase distribution to a single ordinary 
differential equation for the order parameter, yielding an exact 
Poisson kernel representation that is amenable to closed-form 
analytical treatment. Exploiting this structure, we first focus 
on the all-to-all coupled case and derive an exact closed-form 
expression for the effect of a single adversarial kick on the order 
parameter, from which the mechanism responsible for the amplification 
is identified. We then extend the analysis to networks via the annealed 
network approximation, and validate the analytical predictions against 
numerical simulations for all-to-all, Erd\H{o}s--R\'{e}nyi (ER) \cite{albert2002statistical,Takemoto2012_book} and 
Barab\'{a}si--Albert (BA) networks 
\cite{albert2002statistical,barabasi1999emergence}, 
representative network models also considered in 
Ref.~\cite{nagahama2025adversarial}.

\section{Model}

We consider the Kuramoto model of $N$ globally coupled oscillators,
\begin{equation}
    \frac{\mathrm{d}}{\mathrm{d}t}\theta_i = \omega_i 
    + \frac{K}{N}\sum_{j=1}^{N}\sin(\theta_j - \theta_i),
    \label{eq:kuramoto_all}
\end{equation}
where $\theta_i$ and $\omega_i$ are the phase and natural frequency of 
oscillator $i$, and $K > 0$ is the coupling strength. The $1/N$ 
normalization ensures a well-defined thermodynamic limit and follows 
the standard convention for the Ott--Antonsen reduction introduced in 
Sec.~\ref{sec:oa}. The degree of synchronization is measured by the 
complex order parameter
\begin{equation}
    Re^{i\psi} = \frac{1}{N}\sum_{j=1}^{N}e^{i\theta_j},
    \label{eq:order}
\end{equation}
where $R \in [0,1]$ is the synchronization strength and $\psi$ is the 
mean phase. We draw natural frequencies from a Lorentzian distribution 
with center $\omega_0 = 0$ and half-width $\Delta > 0$, for which the 
critical coupling is $K_c = 2\Delta$ \cite{acebron2005kuramoto}.

Following Ref.~\cite{nagahama2025adversarial}, we introduce a 
gradient-based adversarial perturbation applied at regular intervals 
of duration $\tau$. At each perturbation event, every oscillator phase 
is updated as
\begin{equation}
    \theta_i \leftarrow \theta_i 
    + \epsilon\,\mathrm{sign}[\sin(\psi - \theta_i)],
    \label{eq:kick}
\end{equation}
where the sign factor takes the direction of the gradient of $R$ 
with respect to $\theta_i$, so that $\epsilon > 0$ promotes 
synchronization by pushing all phases toward the mean phase $\psi$, 
and $\epsilon < 0$ suppresses it by driving phases away from $\psi$. 
We refer to each such update as a \textit{kick}.

In Sec.~\ref{sec:network}, we extend the analysis to networks with 
degree heterogeneity, where Eq.~\eqref{eq:kuramoto_all} is replaced by
\begin{equation}
    \frac{\mathrm{d}}{\mathrm{d}t}\theta_i = \omega_i 
    + K\sum_{j=1}^{N}A_{ij}\sin(\theta_j - \theta_i),
    \label{eq:kuramoto_net}
\end{equation}
with $A_{ij}$ the adjacency matrix of an unweighted undirected network. 
For the network case, we adopt the unnormalized convention of 
Ref.~\cite{nagahama2025adversarial} to facilitate direct comparison 
with their numerical results. The kick rule of Eq.~\eqref{eq:kick} 
remains unchanged, with $\psi$ computed from the global order 
parameter, Eq.~\eqref{eq:order}.

\section{Ott--Antonsen reduction and kick formula}
\label{sec:oa}

In the thermodynamic limit $N \to \infty$, the OA ansatz
\cite{ott2008low,ott2009long} reduces the dynamics of
Eq.~\eqref{eq:kuramoto_all} to a single ODE for $R(t)$,
\begin{equation}
    \frac{dR}{dt} = -\Delta R + \frac{K}{2}R(1 - R^2),
    \label{eq:oa}
\end{equation}
and implies that the stationary phase distribution takes the form of a
Poisson kernel,
\begin{equation}
    \rho(\phi; R) = \frac{1}{2\pi}
    \cdot\frac{1 - R^2}{1 - 2R\cos\phi + R^2},
    \label{eq:poisson}
\end{equation}
where $\phi = \theta_i - \psi$ is the phase deviation from the mean phase $\psi$, here treated as a continuous integration variable.
The Poisson kernel admits a 
Fourier series with geometrically decaying coefficients, 
$\rho(\phi;R) = (1/2\pi)(1 + 2\sum_{n=1}^{\infty}R^n\cos n\phi)$, 
which enables exact term-by-term integration and underlies 
the closed-form results derived below and in the Appendices.

We now compute the effect of a single adversarial kick of Eq.~\eqref{eq:kick}
on $R$.
The kick formula is derived under the assumption that the phase
distribution immediately before each kick has relaxed to the Poisson kernel
form of Eq.~\eqref{eq:poisson}, which holds for any state on the OA manifold
during the free-evolution intervals between kicks, and is expected to remain
a good approximation for small $\epsilon$, since each kick produces only a
small perturbation to the distribution; the quantitative agreement with
simulations reported below provides a posteriori support for this assumption.

The kick updates each phase deviation $\phi = \theta_i - \psi$ as
\begin{equation}
    \phi_\mathrm{new} = \phi - \epsilon\,\mathrm{sign}(\sin\phi),
\end{equation}
where the sign follows from $\sin(\psi - \theta_i) = -\sin\phi$,
shifting all oscillators toward $\phi = 0$ (i.e., toward the mean phase $\psi$) for
$\epsilon > 0$. Using the symmetry $\rho(-\phi; R) = \rho(\phi; R)$
and the Fourier identity $\int_{-\pi}^{\pi}\rho(\phi;R)\cos\phi\,d\phi = R$,
the post-kick order parameter is (see Appendix~\ref{app:kick} for
details)
\begin{equation}
    R_\mathrm{new} = R\cos\epsilon + 2\sin\epsilon\cdot S(R),
    \label{eq:kick_formula}
\end{equation}
where $S(R) = \int_0^{\pi}\rho(\phi; R)\sin\phi\,d\phi$.
Expanding $\rho(\phi; R)$ in its Fourier series and evaluating this
integral term by term (Appendix~\ref{app:SR}), we obtain the closed
form
\begin{equation}
    S(R) = \frac{(1-R^2)\operatorname{arctanh}(R)}{\pi R},
    \label{eq:SR}
\end{equation}
verified against numerical integration to a relative error of order $10^{-14}$.
In particular, $S(0) = \lim_{R \to 0}S(R) = 1/\pi$, as follows from
$\operatorname{arctanh}(R) \sim R$ for $R \to 0$.

\section{Analytical results: All-to-all case}
\label{sec:mechanism}

\subsection{Effect of a single kick}

The key to understanding the effect of adversarial perturbations
lies in the behavior of the kick formula of Eq.~\eqref{eq:kick_formula}
at small $R$. Since $S(0) = 1/\pi$, the increment per kick is
\begin{equation}
    \Delta R \equiv R_\mathrm{new} - R
    \approx \frac{2\epsilon}{\pi} - O(\epsilon R),
    \label{eq:deltaR}
\end{equation}
which is an $O(1)$ constant independent of $R$. Even when $R \ll 1$,
a single kick changes $R$ by a finite amount proportional to
$\epsilon$, whereas the coupling term in Eq.~\eqref{eq:oa} vanishes
as $R \to 0$ and cannot lift the system out of the incoherent state
for $K < K_c$.

Three concurring factors amplify this effect into large-scale
synchronization changes. First, the $O(1)$ kick efficiency ensures
that the relative boost $\Delta R/R$ diverges as $R \to 0$. Second,
for $K \lesssim K_c$ the restoring force in Eq.~\eqref{eq:oa} is
$\propto (\Delta - K/2) \approx 0$, so the $R$ accumulated by each
kick persists until the next kick rather than decaying away. Third,
an increase in $R$ strengthens the mean field, which promotes further
synchronization in a positive feedback cascade.
The same three factors act in reverse for $\epsilon < 0$: the constant 
decrement $\Delta R \approx 2\epsilon/\pi < 0$ persists even as $R \to 0$,
driving the system toward the incoherent state regardless of $K$,
as analyzed in detail below. Together, these three factors conspire to 
produce synchronization changes far exceeding what the perturbation 
magnitude $\epsilon$ alone would suggest.

\subsection{Hybrid map and fixed points}

\begin{figure}[tbp]
    \includegraphics[width=0.9\columnwidth]{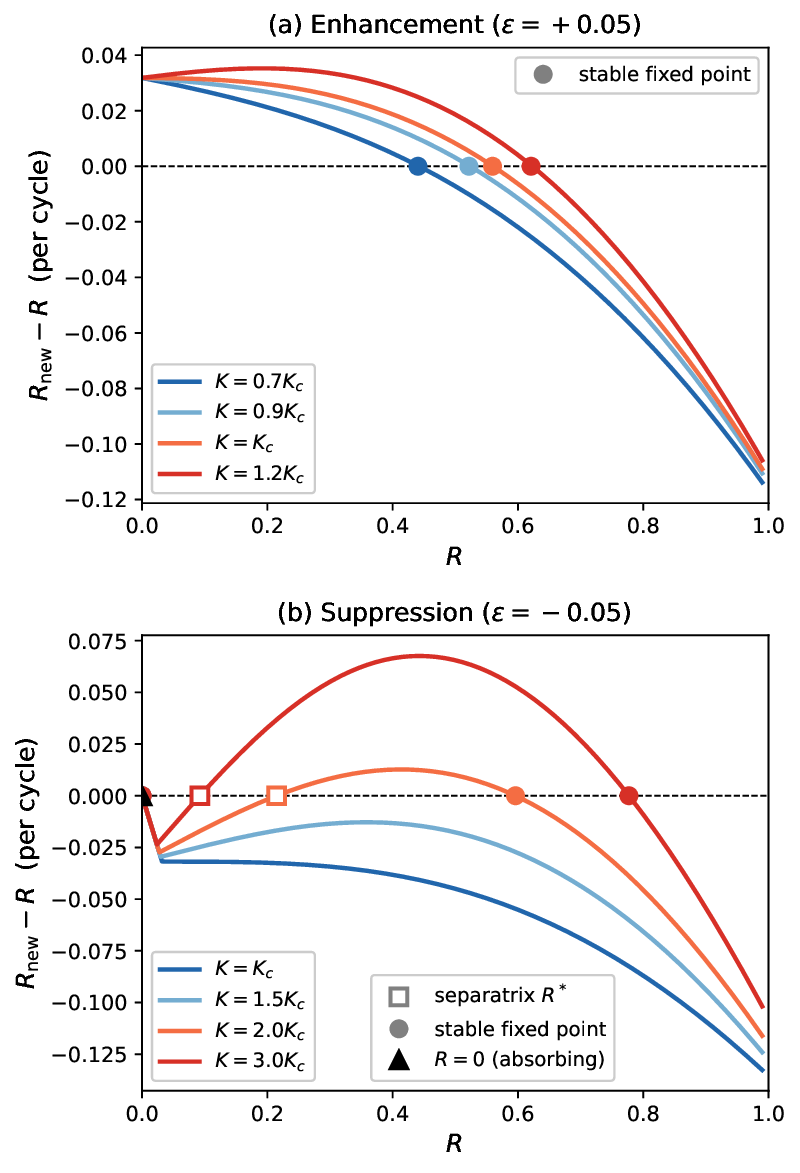}
    \caption{Fixed-point structure of the hybrid map
    $\mathcal{M} = \mathcal{K}_\epsilon \circ \Phi_\tau$
    ($\Delta = 0.5$, $\tau = 0.3$).
    Each curve shows $R_\mathrm{new} - R$ per cycle; fixed points
    occur where curves cross zero.
    (a) Enhancement ($\epsilon = +0.05$): filled circles mark the
    unique stable fixed point.
    (b) Suppression ($\epsilon = -0.05$): open squares mark the
    separatrix $R^*$ and filled circles mark the upper stable fixed
    point. The filled triangle at $R = 0$ indicates the absorbing
    state.}
    \label{fig:fixedpoint}
\end{figure}

The full dynamics is described by a hybrid map $\mathcal{M} =
\mathcal{K}_\epsilon \circ \Phi_\tau$, where $\Phi_\tau$ denotes
integration of Eq.~\eqref{eq:oa} over the interval $\tau$ and
$\mathcal{K}_\epsilon$ is the kick of Eq.~\eqref{eq:kick_formula};
an explicit expression for $\mathcal{M}$ is given in Appendix~\ref{app:phi}.
One application of $\mathcal{M}$, comprising one interval of free
evolution followed by one kick, constitutes a \textit{cycle}.
Steady states correspond to fixed points of $\mathcal{M}$.
Linearizing for $R \ll 1$, the map reads
\begin{equation}
    R_\mathrm{new} \approx \lambda\cos\epsilon\cdot R
    + \frac{2\sin\epsilon}{\pi},
    \quad \lambda \equiv e^{(-\Delta + K/2)\tau}.
    \label{eq:linear_map}
\end{equation}
The fixed-point structure of $\mathcal{M}$ is shown in
Fig.~\ref{fig:fixedpoint}.

For enhancement ($\epsilon > 0$),
the constant term $2\sin\epsilon/\pi > 0$ prevents $R = 0$ from
being a fixed point for any $K$: synchronization is induced at all
coupling strengths and the sharp pitchfork bifurcation at $K_c$ is
replaced by a smooth crossover (Fig.~\ref{fig:fixedpoint}(a)). In
the subcritical regime $K < K_c$, setting $R_\mathrm{new} = R$ in
Eq.~\eqref{eq:linear_map} and solving gives the fixed point
\begin{equation}
    R_\mathrm{ss} \approx \frac{2\sin\epsilon/\pi}
    {1 - \lambda\cos\epsilon},
    \label{eq:Rss}
\end{equation}
valid for $R \ll 1$.
As $K \to K_c$, $\lambda \to 1$ and the denominator 
$1 - \lambda\cos\epsilon$ becomes small, so $R_\mathrm{ss}$
grows large. This is the mathematical origin of
the disproportionate amplification.

For suppression ($\epsilon < 0$),
the constant term $2\sin\epsilon/\pi < 0$ pushes $R$ toward zero
at every kick, competing with the restoring force $\propto R$ from
Eq.~\eqref{eq:oa}. This competition produces bistability: the hybrid
map possesses two stable fixed points, $R = 0$ and an upper
synchronized state, separated by an unstable fixed point
(separatrix) obtained by the same fixed-point condition
$R_\mathrm{new} = R$ applied to Eq.~\eqref{eq:linear_map}, yielding
\begin{equation}
    R^* = \frac{2|\sin\epsilon|}{\pi(\lambda\cos\epsilon - 1)},
    \label{eq:separatrix}
\end{equation}
valid for $R^* \ll 1$, which exists when $\lambda\cos\epsilon > 1$
(Fig.~\ref{fig:fixedpoint}(b)). Since $\lambda = e^{(-\Delta+K/2)\tau}$,
this condition reduces to $K \gtrsim K_c$ for small $|\epsilon|$,
meaning that bistability requires supercritical coupling; for
$K < K_c$ the bistability disappears and $R = 0$ is the unique
attractor.
For random initial conditions $R \sim 1/\sqrt{N}$ (which follows from 
the central-limit-theorem scaling of the order parameter for uniformly 
distributed random initial phases), one has $R < R^*$ throughout the 
practically relevant parameter range, so that $R = 0$ is the only 
reachable steady state in the deterministic limit $N \to \infty$. 
For initial conditions with $R > R^*$, the upper stable fixed point 
is accessible; its value ranges from approximately $0.48$ to $0.79$ 
for the parameters considered here ($\epsilon = -0.03, -0.05, -0.07$, 
$K/K_c = 2$ and $3$), consistent with the finite $R$ values observed 
at large $K$ in Fig.~\ref{fig:transition} for $\epsilon < 0$.

\subsection{Comparison with simulations}
\label{sec:alltoall_comparison}

\begin{figure}[tbp]
    \includegraphics[width=\columnwidth]{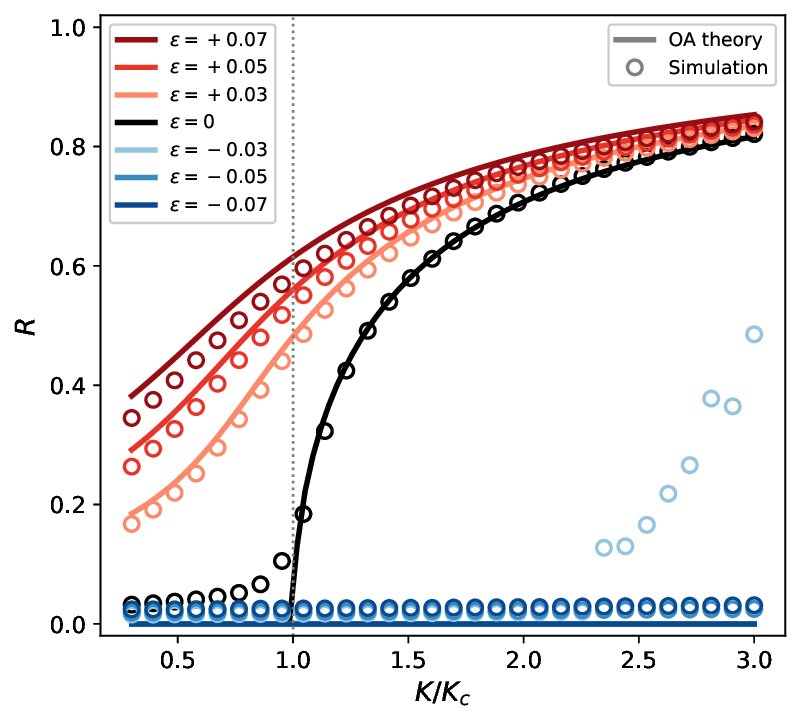}
    \caption{Synchronization transition curves $R$ vs $K$ for the
    all-to-all coupled model ($\Delta = 0.5$, $\tau = 0.3$,
    $N = 1000$). Solid lines: OA theory; symbols:
    simulations averaged over
    100 independent runs. Colors indicate $\epsilon = 0$ (black),
    $\epsilon = +0.03, +0.05, +0.07$ (red tones), and
    $\epsilon = -0.03, -0.05, -0.07$ (blue tones).}
    \label{fig:transition}
\end{figure}

The analytical predictions are tested against numerical simulations
of Eq.~\eqref{eq:kuramoto_all} with Lorentzian frequency
distribution ($\Delta = 0.5$), perturbation interval $\tau = 0.3$,
and $N = 1000$, averaged over 100 independent runs. Figure~\ref{fig:transition}
shows $R$ as a function of $K$ for several values of $\epsilon$.

For $\epsilon > 0$, the OA theory is in quantitative agreement with
simulations: the leftward shift of the transition curve and its
$\epsilon$ dependence are accurately reproduced by
Eq.~\eqref{eq:Rss}. For $\epsilon < 0$, the theory predicts
$R_\mathrm{ss} = 0$ for all $K$ in the deterministic limit, whereas
simulations show a finite $R$ at large $K$. This discrepancy is
attributable to noise-induced escape over the separatrix of
Eq.~\eqref{eq:separatrix} driven by finite-$N$ fluctuations.
When $K$ is sufficiently large and $R^*$ is sufficiently small,
fluctuations of order $1/\sqrt{N}$ can drive $R$ above $R^*$ and
into the basin of the upper stable fixed point.

\begin{figure}
    \includegraphics[width=\columnwidth]{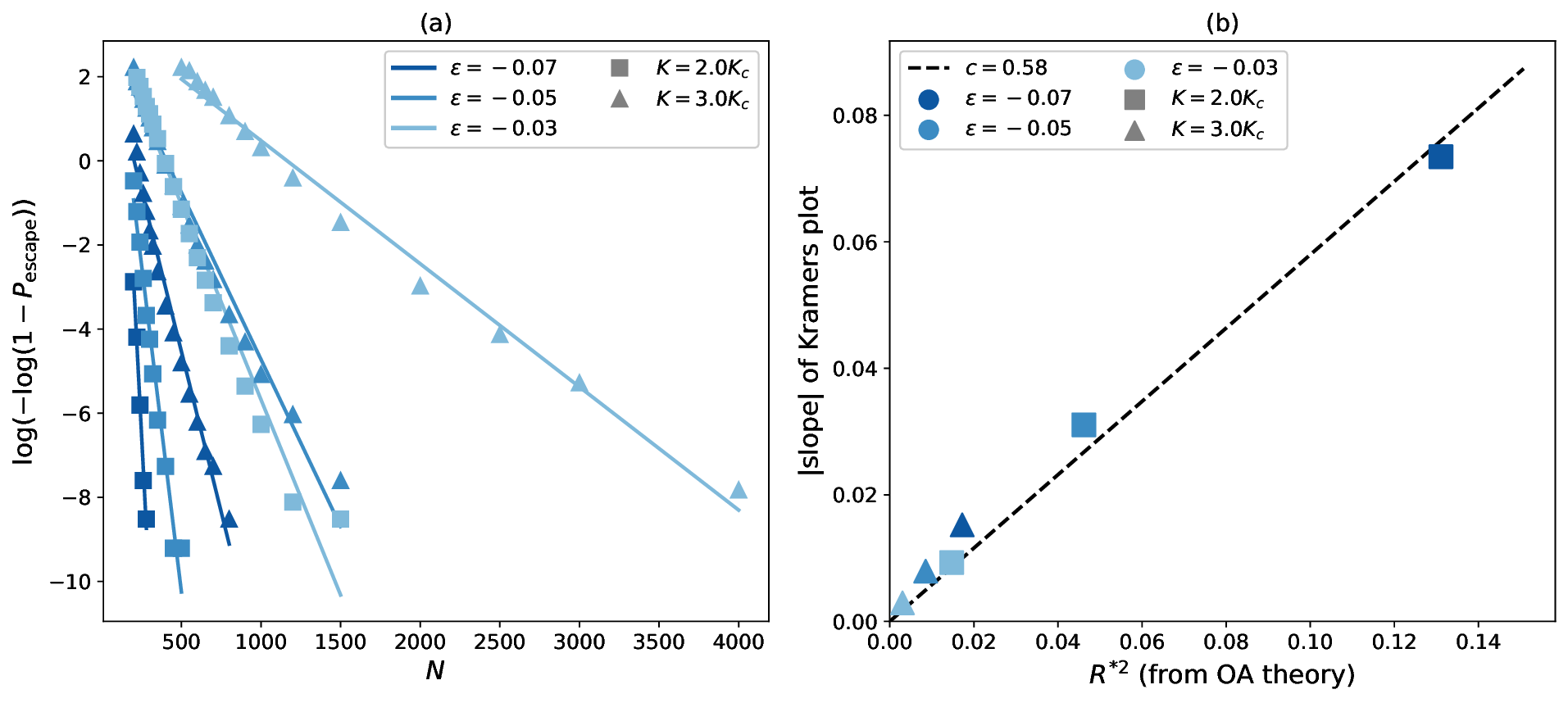}
    \caption{Verification of the Kramers escape picture for the
    suppression side ($\epsilon < 0$, all-to-all coupling,
    $\Delta = 0.5$, $\tau = 0.3$).
    Each data point is estimated from 10000 independent runs.
    (a) $\log(-\log(1 - P_\mathrm{escape}))$ as a function of $N$
    for $\epsilon = -0.03, -0.05, -0.07$ and $K/K_c = 2.0$
    and $3.0$. Filled symbols are simulation data;
    solid lines are linear fits.
    (b) Absolute slopes of the linear fits in (a) plotted against
    $R^{*2}$ computed from Eq.~\eqref{eq:separatrix}. The dashed
    line is a linear fit through the origin with slope $c \approx
    0.58$.}
    \label{fig:kramers}
\end{figure}

To quantify this picture, we measure the escape probability
$P_\mathrm{escape}$, defined operationally as the fraction of
independent runs in which $R$ reaches or exceeds $0.05$
(chosen to exceed the separatrix $R^*$ in all conditions
considered), as a function of $N$ (Fig.~\ref{fig:kramers}(a)).
In a system of $N$ oscillators, finite-size fluctuations of the
order parameter act as an effective noise of intensity
$\sigma^2 \sim 1/N$; assuming a locally quadratic potential
barrier near $R = 0$, the barrier height is proportional to
$R^{*2}$, and Kramers escape
theory~\cite{kramers1940brownian,hanggi1990reaction} predicts an
escape rate $r_\mathrm{esc} \sim e^{-cR^{*2}N}$, so that
$\log(-\log(1 - P_\mathrm{escape}))$ is linear in $N$ with slope
$-cR^{*2}$, as confirmed in Fig.~\ref{fig:kramers}(a):
\begin{equation}
    \log(-\log(1 - P_\mathrm{escape})) \approx -c R^{*2} N + \mathrm{const}.
    \label{eq:kramers}
\end{equation}
Furthermore, the slopes of these linear fits are proportional to
$R^{*2}$ computed from Eq.~\eqref{eq:separatrix}, with a
proportionality constant $c \approx 0.58$ that is consistent across
different values of $\epsilon$ and $K/K_c$
(Fig.~\ref{fig:kramers}(b)). This provides quantitative support for
the bistability picture and the separatrix location predicted by the
OA theory. The discrepancy between the deterministic OA theory and
finite-$N$ simulations on the suppression side thus reflects a
genuine physical effect, namely noise-induced escape, rather than a
failure of the analytical framework. Although
Ref.~\cite{nagahama2025adversarial} used a Gaussian frequency
distribution, the qualitative conclusions are not expected to depend
on this choice, as the kick formula of Eq.~\eqref{eq:kick_formula}
depends only on the Poisson kernel structure of the phase
distribution.

\section{Extension to networks}
\label{sec:network}

\subsection{Annealed network approximation}
\label{sec:annealed}

We now extend the analysis to networks with arbitrary degree
distributions via the annealed network
approximation~\cite{ichinomiya2004frequency,restrepo2005onset},
in which the adjacency matrix is replaced by its ensemble average.
For an uncorrelated random network with degree distribution $P(k)$
and mean degree $\langle k\rangle$, oscillators sharing the same
degree $k$ are then statistically equivalent~\cite{hong2013link,um2014nature}, and the OA ansatz can be applied independently to each degree
class, since oscillators within
the same degree class experience the same mean-field coupling and
their phase distribution consequently follows the Poisson kernel
form of Eq.~\eqref{eq:poisson}. Both ER and BA networks are
structurally uncorrelated in the large-$N$ limit and therefore
fall within the scope of this approximation.

Let $R_k(t)$ denote the local order parameter of degree-$k$
oscillators. The global mean field is~\cite{ichinomiya2004frequency,restrepo2005onset}
\begin{equation}
    H = \sum_k \frac{k\,P(k)}{\langle k\rangle}\,R_k,
    \label{eq:meanfield}
\end{equation}
and the OA equation for each degree class reads~\cite{yoon2015critical}
\begin{equation}
    \frac{dR_k}{dt} = -\Delta R_k
    + \frac{kK}{2}\,H\,(1 - R_k^2).
    \label{eq:Rk_ode}
\end{equation}
The synchronization onset occurs at
\begin{equation}
    K_c = \frac{2\Delta\langle k\rangle}{\langle k^2\rangle},
    \label{eq:Kc}
\end{equation}
which reduces to $K_c = 2\Delta$ in the all-to-all
limit~\cite{ichinomiya2004frequency,restrepo2005onset}.

\subsection{Kick formula and fixed-point conditions}
\label{sec:kick_network}

Because the kick formula~\eqref{eq:kick_formula} was derived from
the Poisson kernel representation of any sub-population obeying the
OA ansatz, and each degree class independently follows the OA
dynamics via Eq.~\eqref{eq:Rk_ode}, the kick applies to each
degree class independently.  Immediately after a kick of strength
$\epsilon$,
\begin{equation}
    R_{k,\mathrm{new}} = R_k\cos\epsilon
    + 2\sin\epsilon\cdot S(R_k),
    \label{eq:kick_network}
\end{equation}
with $S(\cdot)$ as defined in Eq.~\eqref{eq:SR}.  The hybrid map
$\mathcal{M} = \mathcal{K}_\epsilon \circ \Phi_\tau$ therefore
acts on the vector $(R_k)$: the free-flow $\Phi_\tau$ evolves each
$R_k$ according to Eq.~\eqref{eq:Rk_ode} for duration $\tau$, and
$\mathcal{K}_\epsilon$ then applies Eq.~\eqref{eq:kick_network}
componentwise.

The fixed-point condition $\mathcal{M}(R_k^*) = R_k^*$ must be
satisfied self-consistently with Eq.~\eqref{eq:meanfield}.
Setting $dR_k/dt = 0$ in Eq.~\eqref{eq:Rk_ode} and solving for $R_k^*$ gives
\begin{equation}
    R_k^* = \frac{-\Delta + \sqrt{\Delta^2 + (kKH^*)^2}}{kKH^*},
    \label{eq:Rk_fp}
\end{equation}
for $kKH^* > 0$, where $H^*$ is determined by substituting this into Eq.~\eqref{eq:meanfield}.
With kicks present, the steady-state $R_k^*$ is obtained by
solving the system formed by Eqs.~\eqref{eq:kick_network}
and~\eqref{eq:meanfield} numerically for given $K$, $\Delta$,
$\epsilon$, and $P(k)$.

\subsection{Comparison with simulations}
\label{sec:network_sim}

\begin{figure}
    \includegraphics[width=\columnwidth]{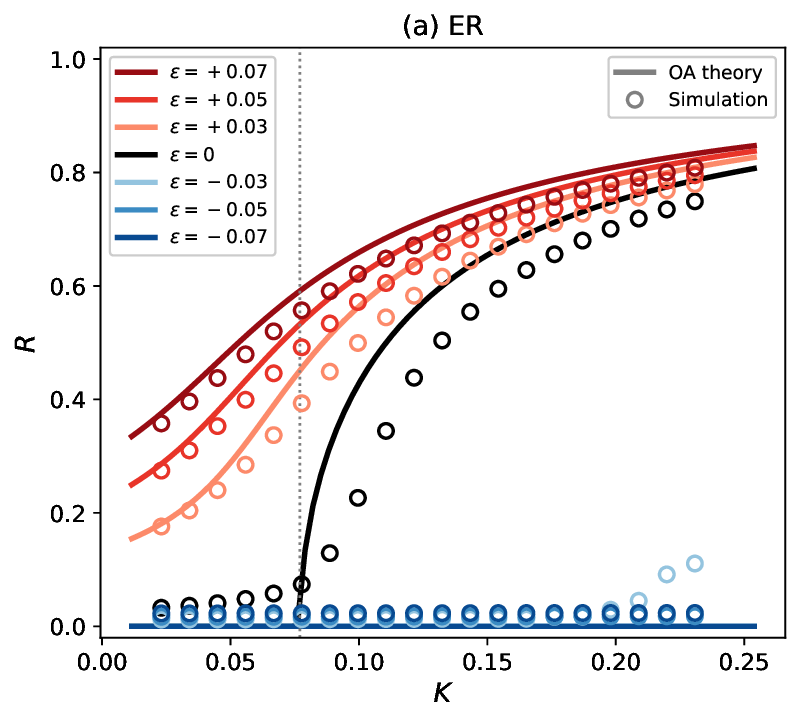}\\
    \includegraphics[width=\columnwidth]{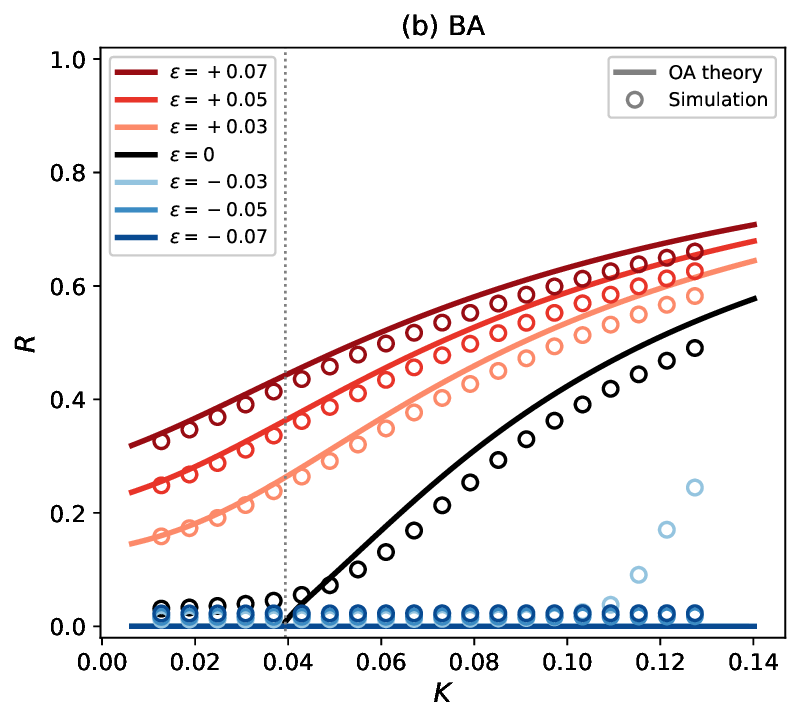}
    \caption{Synchronization transition curves $R$ vs.\ $K$ for
    (a) ER and (b) BA networks ($N = 1000$,
    $\langle k\rangle = 12$, $\Delta = 0.5$, $\tau = 0.3$).  Solid lines: OA theory; symbols: simulations averaged over 100 independent realizations.
    Colors indicate $\epsilon = 0$ (black),
    $\epsilon = +0.03, +0.05, +0.07$ (red tones), and
    $\epsilon = -0.03, -0.05, -0.07$ (blue tones).
    The dotted vertical line marks $K_c$ from
    Eq.~\eqref{eq:Kc}.}
    \label{fig:network}
\end{figure}

Figure~\ref{fig:network} shows $R$ vs.\ $K$ for ER and BA networks
($\langle k\rangle = 12$, $N$ and other parameters as in
Sec.~\ref{sec:mechanism}) under periodic kicks.
ER networks have a Poissonian degree distribution with modest
fluctuations \cite{albert2002statistical,Takemoto2012_book}, while BA networks are scale-free with a power-law
tail ($P(k) \sim k^{-3}$)~\cite{albert2002statistical,barabasi1999emergence} and relatively strong degree heterogeneity.
For the OA theory curves, the degree distribution of ER networks
is approximated by a Poisson distribution with mean $\langle
k\rangle$, while for BA networks the degree moments $\langle
k\rangle$ and $\langle k^2\rangle$ are estimated by averaging
over multiple network realizations of size $N$, from which $K_c$
is computed via Eq.~\eqref{eq:Kc}.
The annealed OA theory (solid curves) is compared with
direct numerical simulation of Eq.~\eqref{eq:kuramoto_net}
(symbols) for both signs of $\epsilon$.

The theory captures the overall behavior for both network types:
the ordering of transition curves with $\epsilon$, the leftward
shift for $\epsilon > 0$, and the suppression of $R$ for
$\epsilon < 0$ are all reproduced quantitatively for BA and
qualitatively for ER networks. For sparse quenched networks,
however, the annealed approximation is known to underestimate the
effective disorder, shifting the apparent transition to larger $K$
and broadening the transition curve~\cite{hong2013link,um2014nature,gkogkas2022graphop}.
This is visible for ER networks, where the simulated transition is
systematically shifted to larger $K$ relative to the theoretical
prediction. For BA networks the agreement appears closer, despite
the broader degree distribution: the very small $K_c$ predicted by
Eq.~\eqref{eq:Kc} places the transition in a regime where $R$
grows gradually with $K$, making deviations less pronounced.
In both cases we have confirmed numerically that the agreement 
improves with increasing $N$, consistent with the annealed 
approximation becoming exact in the thermodynamic 
limit~\cite{restrepo2005onset}.
For $\epsilon < 0$, a small number of simulation points show finite
$R$ at large $K$, attributable to noise-induced escape over the
separatrix as discussed in Sec.~\ref{sec:alltoall_comparison}.

Watts--Strogatz networks, also examined in
Ref.~\cite{nagahama2025adversarial}, are not addressed by the
present framework: the annealed approximation is inapplicable when
clustering is strong, because correlated local environments
invalidate the degree-class factorization of
Eq.~\eqref{eq:Rk_ode}.

\subsection{Hub and periphery roles}
\label{sec:hub}

\begin{figure}
    \includegraphics[width=0.9\columnwidth]{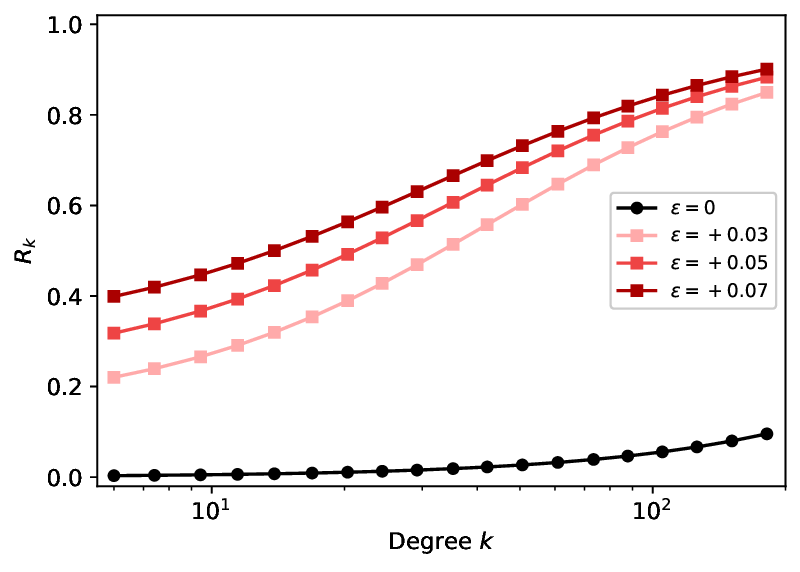}
    \caption{Degree-resolved local order parameter $R_k$ for a BA
    network at $K = K_c$, computed from the annealed OA theory
    ($\langle k\rangle = 12$, $\Delta = 0.5$, $\tau = 0.3$).
    Colors indicate $\epsilon = 0$ (black) and
    $\epsilon = +0.03, +0.05, +0.07$ (red tones).}
    \label{fig:Rk}
\end{figure}

The broad degree distribution of BA networks stratifies $R_k^*$
strongly across degree classes, as shown in Fig.~\ref{fig:Rk}.
At $K = K_c$ without kicks ($\epsilon = 0$), $R_k^* \approx 0$
for all degrees, as expected at the critical point.  Enhancement
kicks induce a degree-dependent profile: Eq.~\eqref{eq:Rk_fp}
shows that $R_k^*$ increases monotonically with $k$ for fixed
$H^* > 0$, since a larger $k$ strengthens the effective mean-field
coupling $kKH^*$, so that hub nodes reach large $R_k^*$ while
low-degree nodes remain at smaller values.

This stratification has two competing consequences for the kick
dynamics.  On one hand, low-degree nodes operate at small $R_k^*$,
where $S(R_k^*)$ is appreciable (cf.\ Eq.~\eqref{eq:SR}), and
therefore experience a substantial kick-induced increment per cycle.
On the other hand, because $S(R) \to 0$ as $R \to 1$, hubs with
large $R_k^*$ are nearly insensitive to kicks; yet hubs dominate
the mean field $H^*$ through the $k$-weight in
Eq.~\eqref{eq:meanfield}, while low-degree nodes contribute little
to $H^*$ despite their high kick sensitivity.  Kick sensitivity is
thus decoupled from mean-field dominance: the nodes most responsive
to kicks exert the least influence on $H^*$, weakening the positive
feedback loop and explaining why the same $\epsilon$ produces a
smaller global effect in BA than in ER networks.

For suppression ($\epsilon < 0$), the same decoupling makes BA
networks harder to desynchronize.  Hubs sustain large $R_k^*$ and
continue to dominate $H^*$ even as peripheral nodes are driven
toward incoherence, because kicks are least effective precisely
where the mean-field contribution is greatest.  In the deterministic
limit $N \to \infty$, global desynchronization requires $|\epsilon|$
large enough that the self-consistency equations
Eqs.~\eqref{eq:kick_network} and~\eqref{eq:meanfield} admit no
solution with $H^* > 0$.  This condition becomes increasingly
difficult to satisfy as $\langle k^2\rangle/\langle k\rangle$
grows, and is therefore harder to meet for BA than for ER networks
of the same mean degree.

\section{Conclusion}
\label{sec:conclusion}

We have provided an analytical foundation for adversarial
synchronization control in Kuramoto networks using the
Ott--Antonsen reduction.  For the all-to-all coupled case, we
derived an exact closed-form kick formula whose key quantity,
$S(0) = 1/\pi$, produces an $O(1)$ increment in $R$ per kick
even at $R = 0$, independent of the coupling strength.  Combined
with slow relaxation near $K_c$ and mean-field feedback, this
explains the disproportionate amplification reported in
Ref.~\cite{nagahama2025adversarial}.  The fixed-point analysis
further reveals a fundamental asymmetry between enhancement and
suppression, with the latter governed by noise-induced escape in
finite systems, implying that perfect suppression is achievable
only in the thermodynamic limit $N \to \infty$.  Extending the
framework to degree-heterogeneous networks via the annealed
approximation, we showed that the theory captures the overall
synchronization behavior of both ER and BA networks, and
identified a decoupling between kick sensitivity and mean-field
dominance in BA networks that limits the effectiveness of uniform
adversarial control in scale-free topologies.

Several extensions remain for future work.  A systematic treatment
of finite-$N$ fluctuations, which govern noise-induced escape on
the suppression side, could be pursued via the circular cumulant
framework~\cite{tyulkina2018dynamics,goldobin2021reduction}, which
extends the OA reduction to noisy oscillator populations.  The
present analysis assumes a Lorentzian frequency distribution; 
extending the framework to other distributions, such as the 
Gaussian distribution used in Ref.~\cite{nagahama2025adversarial}, 
would broaden its applicability, though the qualitative conclusions 
are not expected to change.  The annealed
approximation adopted here breaks down for correlated network
topologies; extensions of the OA framework to assortative and
disassortative networks have been developed~\cite{restrepo2014mean}
and could serve as a starting point for treating such cases.

Despite these limitations, the analytical framework introduced here
provides a tractable foundation for understanding and designing
kick-based synchronization control, and is likely applicable
beyond the specific gradient-based rule of
Ref.~\cite{nagahama2025adversarial} to other impulsive
perturbation strategies in oscillator networks.


%
%

%

\begin{acknowledgments}
This study was funded by JSPS KAKENHI (grant number 26K03029).
\end{acknowledgments}

\section*{Data Availability}
The code used in this study is publicly available at \url{https://github.com/kztakemoto/advSyncOA}.

\bibliography{references}

\appendix

\section{Derivation of the Kick Formula}
\label{app:kick}

We derive Eq.~\eqref{eq:kick_formula}. After the kick, the order
parameter becomes
\begin{equation}
    R_\mathrm{new} = \left|\int_{-\pi}^{\pi}
    \rho(\phi; R)\,e^{i\phi_\mathrm{new}}\,d\phi\right|.
\end{equation}
Using $\phi_\mathrm{new} = \phi - \epsilon\,\mathrm{sign}(\sin\phi)$
and the symmetry $\rho(-\phi;R) = \rho(\phi;R)$, the imaginary part
of the integral vanishes and we obtain
\begin{align}
    R_\mathrm{new}
    &= 2\int_0^{\pi}\rho(\phi; R)\cos(\phi - \epsilon)\,d\phi
    \nonumber\\
    &= 2\cos\epsilon\int_0^{\pi}\rho(\phi;R)\cos\phi\,d\phi \nonumber\\
    &~~~~~~ + 2\sin\epsilon\int_0^{\pi}\rho(\phi;R)\sin\phi\,d\phi.
\end{align}
The first integral evaluates to $R/2$ by the Fourier property of the
Poisson kernel, $\int_{-\pi}^{\pi}\rho(\phi;R)\cos\phi\,d\phi = R$,
and the second integral defines $S(R)$ as given in Sec.~\ref{sec:oa}.
Substituting yields Eq.~\eqref{eq:kick_formula}.

\section{Closed-Form Evaluation of $S(R)$}
\label{app:SR}

We evaluate $S(R) = \int_0^{\pi}\rho(\phi;R)\sin\phi\,d\phi$ using
the Fourier expansion of the Poisson kernel,
\begin{equation}
    \rho(\phi; R) = \frac{1}{2\pi}
    \left(1 + 2\sum_{n=1}^{\infty}R^n\cos n\phi\right).
\end{equation}
Since $\int_0^{\pi}\cos(n\phi)\sin\phi\,d\phi$ vanishes for odd $n$
and equals $-2/(n^2 - 1)$ for even $n = 2m$, we obtain
\begin{equation}
    S(R) = \frac{1}{\pi} - \frac{2}{\pi}
    \sum_{m=1}^{\infty}\frac{R^{2m}}{4m^2 - 1}.
\end{equation}
Applying the partial fraction decomposition
$1/(4m^2-1) = \frac{1}{2}(1/(2m-1) - 1/(2m+1))$ and the series
expansion $\operatorname{arctanh}(R) = \sum_{m=0}^{\infty}R^{2m+1}/(2m+1)$,
one finds
\begin{equation}
    \sum_{m=1}^{\infty}\frac{R^{2m}}{4m^2 - 1}
    = \frac{R - (1-R^2)\operatorname{arctanh}(R)}{2R}.
\end{equation}
Substituting and simplifying yields Eq.~\eqref{eq:SR}.

\section{Analytical Solution of the OA Flow}
\label{app:phi}

The OA equation, Eq.~\eqref{eq:oa}, is separable. Setting $u = R^2$, it becomes
\begin{equation}
    \frac{du}{dt} = \beta u - K u^2, \quad \beta = K - 2\Delta,
\end{equation}
which is a logistic equation with exact solution \cite{ott2008low}
\begin{equation}
    u(\tau) = \frac{\beta u_0 e^{\beta\tau}}
    {\beta + K u_0(e^{\beta\tau}-1)},
\end{equation}
where $u_0 = R^2$. This gives
\begin{equation}
    \Phi_\tau(R) = \sqrt{\frac{\beta R^2 e^{\beta\tau}}
    {\beta + KR^2(e^{\beta\tau}-1)}},
    \label{eq:phi_tau}
\end{equation}
where $\beta = K - 2\Delta$. At the critical point $K = K_c = 2\Delta$ 
($\beta = 0$), Eq.~\eqref{eq:phi_tau} reduces to
\begin{equation}
    \Phi_\tau(R)\big|_{\beta=0} = \frac{R}{\sqrt{1 + KR^2\tau}}.
\end{equation}
Setting $R' = \Phi_\tau(R)$, the hybrid map 
$\mathcal{M} = \mathcal{K}_\epsilon \circ \Phi_\tau$ then reads
\begin{equation}
    \mathcal{M}(R) = R'\cos\epsilon 
    + 2\sin\epsilon\cdot S(R'),
    \label{eq:hybrid_map}
\end{equation}
with $S(\cdot)$ as defined in Eq.~\eqref{eq:SR}.

\end{document}